\documentclass[conference]{IEEEtran}

\usepackage{fancyhdr, epsfig, epsf, amsthm, amsmath, amssymb, amsfonts, subfigure,color}
\usepackage{threeparttable}
\usepackage[noadjust]{cite}
\usepackage{dsfont}
\usepackage{enumerate}
\usepackage{comment}

\newtheorem{lemma}{Lemma}
\newtheorem{corollary}{Corollary}

\newtheorem{proposition}{Proposition}

\def\E{\mathsf{E}}

\def\SNR{\mathsf{SNR}}
\def\SIR{\mathsf{SIR}}

\def\({\left(}
\def\){\right)}

\def\[{\left[}
\def\]{\right]}

\usepackage{multirow}
\usepackage{bigstrut}

\def\SIRn{\widetilde{\SIR}}

\def\R{{\hat{R}_1}}
\def\OMA{\text{O}}
\def\NOMA{\text{N}}
\def\L{\text{Link}}
\def\SIRn{\overline{\SIR}}
\def\thmspacing{1.1}
\def\figwidth{8.5 cm}

\def\papertitle{URLLC-eMBB Slicing to Support VR Multimodal Perceptions over Wireless Cellular Systems}

\IEEEoverridecommandlockouts

\begin{document}
\title{ \fontsize{24}{28}\selectfont  \papertitle}

\author{Jihong~Park and Mehdi~Bennis
\thanks{J.~Park and M.~Bennis are with the Centre for Wireless Communications, University of Oulu, Oulu 90014, Finland (email: \{jihong.park, mehdi.bennis\}@oulu.fi). }
}

\maketitle \thispagestyle{empty}

\begin{abstract} Virtual reality (VR) enables mobile wireless users to experience multimodal perceptions in a virtual space. In this paper we investigate the problem of concurrent support of visual and haptic perceptions over wireless cellular networks, with a focus on the downlink transmission phase. While the visual perception requires moderate reliability and maximized rate, the haptic perception requires fixed rate and high reliability. Hence, the visuo-haptic VR traffic necessitates the use of two different network slices: enhanced mobile broadband (eMBB) for visual perception and ultra-reliable and low latency communication (URLLC) for haptic perception. We investigate two methods by which these two slices share the downlink resources orthogonally and non-orthogonally, respectively. We compare these methods in terms of the just-noticeable difference (JND), an established measure in psychophysics, and show that non-orthogonal slicing becomes preferable under a higher target integrated-perceptual resolution and/or a higher target rate for haptic perceptions.
\end{abstract}

\section{Introduction}
Virtual reality (VR) is often seen as one of the most important applications in 5G cellular systems~\cite{ABIQualcommVR:17,EjderVR:17,ParkWCL:18}. As in real life, mobile VR users can interact in the virtual space immersively with virtual objects that may stimulate their multiple sensory organs. This multimodal VR perception happens, for example, when a VR user measures the size of a virtual object through visual and haptic senses. In this study, we consider the problem of supporting such visuo-haptic VR perceptions over wireless cellular networks, and focus on the downlink design.

The key challenge is that these two perceptions have completely different cellular service requirements. In fact, visual traffic requires high data rate and relatively low reliability with packet error rate (PER) on the order of $10^{-1}\sim 10^{-3}$~\cite{Shi:10,UR2Cspaswin:17}. This requirements can be supported mostly through enhanced mobile broadband (eMBB) links~\cite{ITU5G:15}. Haptic traffic, by contrast, should guarantee a fixed target rate and high reliability with PER on the order of $10^{-4}\sim 10^{-5}$~\cite{Steinbach:12,Zhang:18}, which can be satisfied via ultra-reliable and low latency communication (URLLC) links~\cite{PetarURLLC:17,MehdiURLLC:18}.

Furthermore, in order to render a smooth multimodal experience, the PERs associated with the visuo-haptic VR perceptions should guarantee a target perceptual resolution. To be precise, the perceptual resolution is commonly measured by using the just-noticeable difference (JND) in psychophysics, a field of study that focuses on the quantitative relation between physical stimulus and perception \cite{Ernst:2002aa,Shi:10,ShiHirche:16}. Following Weber's law, JND describes the minimum detectable change amount of perceptual inputs, e.g., $3$ mm for the object size measurement using visuo-haptic perceptions~\cite{Ernst:2002aa}. According to psychophysical experiments, the JND of the aggregate visuo-haptic perception is the harmonic mean of the squared JNDs of the individual perceptions~\cite{Ernst:2002aa}, in which the JND of each perception is proportional to the PER~\cite{Shi:10}.

\begin{figure}%
\centering
\includegraphics[width=  8.5 cm ]{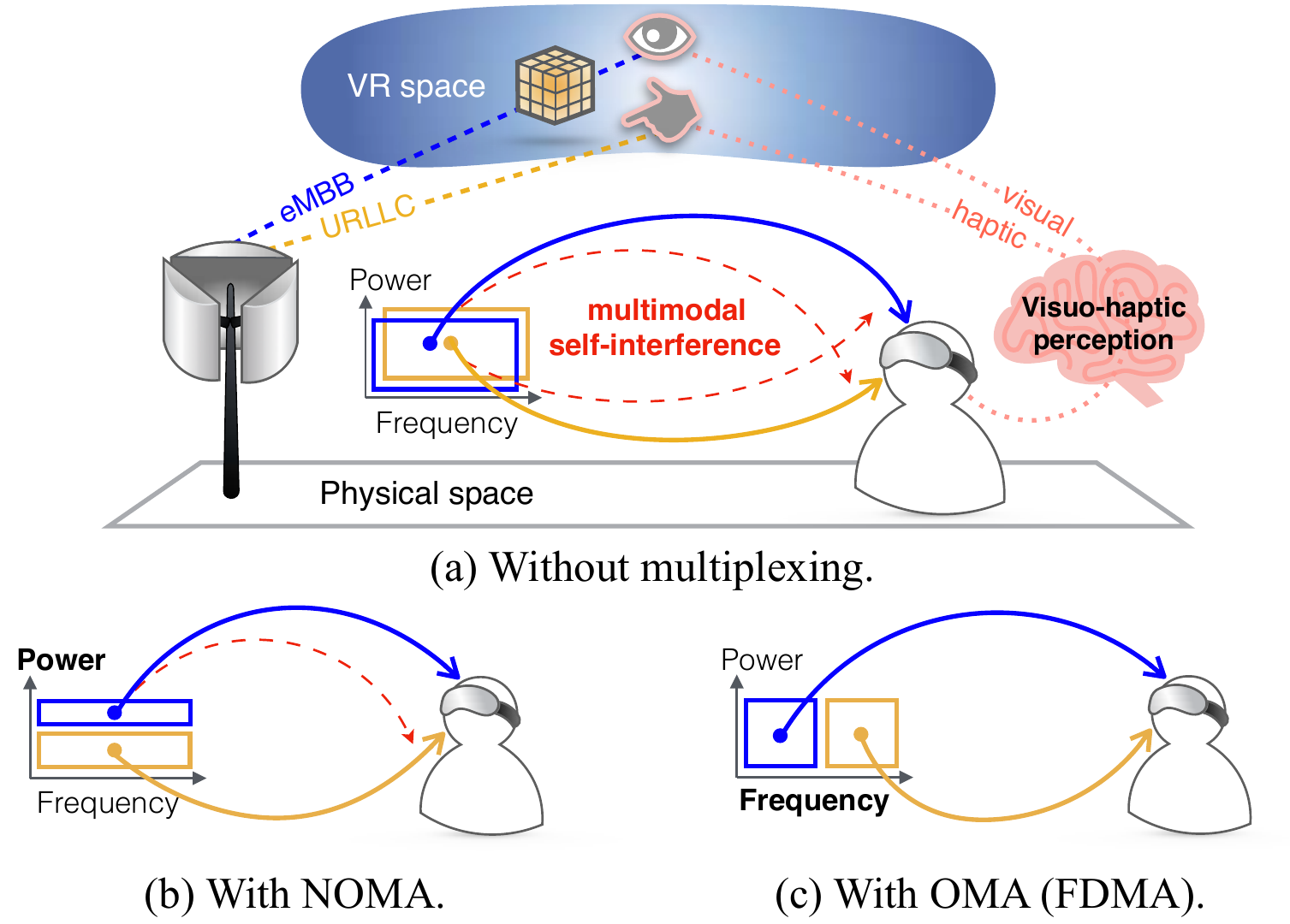}
\caption{\small{An illustration of visuo-haptic VR perceptions and the required URLLC-eMBB traffic slicing: (a) without multiplexing; (b)~with NOMA; and (c) with OMA.}}
\end{figure}

As a result, the PERs associated with the visuo-haptic VR traffic should be adjusted so as to achieve a target JND, while abiding by the eMBB and URLLC service objectives in terms of PERs and data rates. Due to the discrepancy of the visual and haptic service requirements, it is difficult to support both perceptions through either eMBB or URLLC links. Hence, it is necessary to \emph{slice the visuo-haptic VR traffic into eMBB and URLLC links}, leading to URLLC-eMBB multimodal transmissions. Unfortunately, such multimodal transmissions bring about \emph{multimodal self-interference}, which is manifested through an actual wireless interference as visualized in Figs. 1-a and b, or via the necessity to share resources as shown in Figs.~1-b and~c.

This critical self-interference can be alleviated by multiplexing the URLLC-eMBB multimodal transmissions over the transmit power domain with successive interference cancellation (SIC) at reception, i.e., downlink non-orthogonal multiple access (NOMA)~\cite{3GPPMUST:2015}, as illustrated in Fig.~1-b. Alternatively, as Fig.~1-c shows, the self-interference can be avoided via orthogonal multiple access (OMA) such as frequency division multiple access (FDMA). In this paper, using stochastic geometry, we investigate the optimal design of NOMA and OMA to support visuo-haptic VR perceptions while coping with the multimodal self-interference in a large-scale downlink system.

\textbf{Related Works} -- 
The communication and computation resource management of mobile VR networks has recently been investigated in \cite{EjderVR:17,OsvaldoAR:17,ChenSaad:17,ParkWCL:18,Elbamby:18}, particularly under a VR social network application~\cite{ParkWCL:18} and a VR gaming scenario~\cite{Elbamby:18}. The end-to-end latency has been studied in \cite{OsvaldoAR:17} for a single-cell scenario and in \cite{ChenSaad:17,ParkWCL:18} for a multi-cell scenario. These works focus primarily on supporting either visual or haptic perceptions. Towards supporting multimodal perceptions, suitable network architecture and coding design have been proposed in~\cite{Zhang:18,Steinbach:12}, while not specifying the requirements on the wireless links. In an uplink single-cell system, orthogonal/non-orthogonal multiplexing of URLLC and eMBB links has been optimized by exploiting their reliability~diversity in \cite{Petar5G:18}.


\textbf{Contributions} --
The main contributions of this work are summarized as follows.
\begin{itemize}
\item To the best of our knowledge, this is the first work that combines both visual and haptic modalities in the context of mobile VR network design.

\item To support visuo-haptic VR perceptions, an optimal downlink NOMA design with reliability-ordered SIC has been proposed (see {Lemma 2} and {Proposition~4}).

\item Compared to an OMA baseline (see {Proposition~2}), it has been observed that the proposed NOMA becomes preferable under a higher target integrated-perceptual resolution and/or a higher target rate for haptic perceptions (see {Fig.~3}).

\item By using stochastic geometry, closed-form average rate expressions have been derived for downlink URLLC-eMBB multiplexing under OMA and NOMA in a large-scale cellular network (see {Propositions}~{1} and {3}).

\end{itemize}

\section{System Model and Problem Formulation}
In this section, we first introduce the downlink system operation of OMA and NOMA under a single-cell scenario, and describe its extension to the operation under a large-scale network. Then, we specify visuo-haptic perceptions, followed by the problem formulation of visuo-haptic VR traffic slicing and multiplexing.

The user under study requests visuo-haptic VR perceptions that are supported through URLLC-eMBB cellular links. We use the subscript $i\in\{1,2\}$ to indicate the URLLC link $\L_1$ with $i=1$ and the eMBB link $\L_2$ with $i=2$. The subscript $j\in\{\OMA,\NOMA\}$ identifies OMA and NOMA, respectively. 

\subsection{Single-Cell Channel Model with OMA and NOMA}
In a downlink scenario, we consider a single user that is associated with a single base station (BS). For both OMA and NOMA, the transmissions of the BS at a given time occupy up to the frequency bandwidth normalized to one, which is divided into the $K$ number of miniblocks. Each miniblock is assumed to be within the frequency-time channel coherence intervals. The channel coefficients are thus constants within each miniblock, and fade independently across different miniblocks over frequency and time. The transmit power of the BS is equally divided for each miniblock, normalized to~one.

\subsubsection{Single-Cell OMA} \label{Sect:Sys_OMA}
A set $\mathcal{K}_i$ of miniblocks are allocated to $\L_i$, with $|\mathcal{K}_1|+|\mathcal{K}_2|=K$. Each set corresponds to a fraction $w_{i,\OMA}=|\mathcal{K}_i|/K>0$, with $w_{1,\OMA}+w_{2,\OMA}=1$. The transmit power allocations to $\L_1$ and $\L_2$ are set as the maximum transmit power per miniblock. Denoting as $\beta_{i,\OMA}\leq 1$ the transmit power allocation fraction to miniblock $k\in\mathcal{K}_i$, this corresponds to the allocations that equal $\beta_{1,\OMA}=\beta_{2,\OMA}=1$.

The user's received signal-to-noise ratio ($\SNR$) is determined by small-scale and large-scale fading gains. For a given user-BS association distance $r$, the large-scale fading gains of $\L_1$ and $\L_2$ are identically given as $r^{-\alpha}$ with the path loss exponent $\alpha>2$. For miniblock $k\in\mathcal{K}_i$, the small-scale fading gain $g_{i}^{(k)}$ is an exponential random variable with unit mean, which is independent and identically distributed (i.i.d.) across different miniblocks. The user's received $\SNR$ of $\L_i$ through miniblock $k\in\mathcal{K}_i$ is then expressed as

\vspace{-10pt}\small\begin{align}
\SNR_{i,\OMA}^{(k)} = \frac{ g_{i}^{(k)} r^{-\alpha}}{ N},\label{Eq:SNR_OMA}
\end{align}\normalsize
where $N$ is the noise spectral density of a single miniblock.

\subsubsection{Single-Cell NOMA} \label{Sect:Sys_NOMA}
The entire bandwidth is utilized for both links in NOMA, i.e., the miniblock allocation fractions equal $w_{1,\NOMA}=w_{2,\NOMA}=1$. This is enabled by transmitting the superposition of the signals intended for $\L_1$ and $\L_2$, with their different transmit power allocations, and then by decoding the signals with SIC at reception~\cite{3GPPMUST:2015}. The transmit power allocated to $\L_i$ has a fraction $\beta_{i,\NOMA}$ of the maximum transmit power per miniblock, with $\beta_{1,\NOMA}+\beta_{2,\NOMA}=1$.

At reception, unless otherwise noted, we consider $\L_1$ is decoded prior to $\L_2$. This SIC order implicitly captures the low-latency guarantee of $\L_1$, as addressed in \cite{Petar5G:18} for an uplink scenario. Furthermore, it improves the overall NOMA system performance due to the reliability diversity of $\L_1$ and $\L_2$, to be elaborated in Sect.~\ref{Sect:OptNOMA}.

With the said SIC order, the signal intended for $\L_1$ is first decoded, while treating the signal for $\L_2$ as noise, i.e., multimodal self-interference. The decoded signal is then removed by applying SIC, and the remaining signal for $\L_2$ is finally decoded without self-interference. The user's received $\SNR$ for $\L_i$ through miniblock $k\in\mathcal{K}_i$ is thereby obtained~as

\vspace{-10pt}\small\begin{align}
\SNR_{1,\NOMA}^{(k)} &= \frac{\beta_{1,\NOMA}   g_{i}^{(k)} r^{-\alpha}}{\beta_{2,\NOMA}  g_{i}^{(k)} r^{-\alpha} + N} \quad\text{and}\label{Eq:SNR1_NOMA}\\ 
\SNR_{2,\NOMA}^{(k)} &= \frac{\beta_{2,\NOMA}   g_{i}^{(k)} r^{-\alpha}}{N}, \label{Eq:SNR2_NOMA}
\end{align}\normalsize
Note that all the fading gains of $\L_1$ and $\L_2$ are identically $g_{i}^{(k)} r^{-\alpha}$ since their channels are identical.

\subsection{Channel Model under a Stochastic Geometric Network}
By using stochastic geometry, the aforementioned single-cell operation of OMA and NOMA is extended to a large-scale multi-cell scenario as follows. The BSs under study are deployed in a two-dimensional Euclidean plane, according to a stationary Poisson point process (PPP) $\Phi$ with density~$\lambda$, where the coordinates $x$ of a BS belongs to $\Phi$. Following the single-cell operation, each BS serves a single user through its $\L_1$ and $\L_2$.

The locations of users follow an arbitrary stationary point process. Each user associates with the nearest BS, and downloads the visuo-haptic VR traffic through the $\L_1$ and $\L_2$ of the BS. Following~\cite{Andrews:2011bg}, we focus our analysis on a typical user that is located at the origin and associated with the nearest BS located at position $x_o$ of the plane. This typical user captures the spatially-averaged performance, thanks to Slyvnyak's theorem~\cite{HaenggiSG} and the stationarity of $\Phi$.

In the previous single-cell scenario, interference occurs only from the multimodal self-interference under NOMA, as shown in~\eqref{Eq:SNR1_NOMA}. In addition to such intra-cell self-interference, extension to the stochastic geometric network model induces inter-cell interference. As done in \cite{Andrews:2011bg,JHParkTWC:15,UR2Cspaswin:17}, inter-cell interference is treated as noise, and is assumed to be large such that the maximum noise power $N$ is negligible. In this interference-limited regime, channel quality is measured not by $\SNR$ but by signal-to-interference ratio ($\SIR$), as described next.

The inter-cell interference is measured by the typical user, and comes from the set $\Phi_o=\Phi\backslash\{x_o\}$ of the BSs that are not associated with the typical user. We consider every BS always utilizes the entire bandwidth and the maximum transmit power. The average inter-cell interference per miniblock is thus identically given under both OMA and NOMA. The instantaneous inter-cell interference varies due to small-scale fading. For each miniblock, any interfering link's small-scale fading is independent of the small-scale fading of the typical user's desired $\L_1$ and $\L_2$. 

Under OMA, the typical user's received $\SIR$ of $\L_i$ through miniblock $k\in\mathcal{K}_i$ is thereby given as

\vspace{-10pt}\small\begin{align}
\SIR_{i,\OMA}^{(k)} &= \frac{ g_{i}^{(k)} |x_o|^{-\alpha}}{  \sum_{x\in\Phi_o } g_{x}^{(k)} |x|^{-\alpha} }, \label{Eq:SIR_OMA}
\end{align}\normalsize
where $g_{x}^{(k)}$'s are exponential random variables with unit mean, which are independent of $g_{i}^{(k)}$ and are i.i.d. across  different interfering BSs. Likewise, under NOMA, the typical user's received $\SIR$ of $\L_i$ through miniblock $k\in\mathcal{K}_i$ is expressed~as

\vspace{-10pt}\small\begin{align}
\SIR_{1,N}^{(k)} &= \frac{\beta_1   g^{(k)} |x_o|^{-\alpha}}{\beta_2  g^{(k)} |x_o|^{-\alpha} + \sum_{x\in\Phi_o } g_{x}^{(k)} |x|^{-\alpha} } \quad\text{and} \label{Eq:SIR1_NOMA}\\ 
\SIR_{2,N}^{(k)}&= \frac{\beta_2   g^{(k)} |x_o|^{-\alpha}}{\sum_{x\in\Phi_o } g_{x}^{(k)} |x|^{-\alpha}}. \label{Eq:SIR2_NOMA}
\end{align}\normalsize 

It is noted that all the $\SIR$s under NOMA and OMA are identically distributed across different miniblocks. For the typical user's $\L_1$  and $\L_2$, the large-scale fading gains are identical. Their small-scale fading gains are independent under OMA, but are fully-correlated under NOMA.

\subsection{Average Rate with Decoding Success Guarantee}
In a large-scale downlink cellular system with OMA and NOMA, we derive the typical user's average rate that guarantees a target decoding success probability. Decoding becomes successful when the instantaneous downlink rate exceeds the transmitted coding rate. 

To facilitate tractable analysis, we consider that the instantaneous channel information is not available at each BS. With the channel information at a BS, one can improve the average rate by adjusting the transmit power \cite{Petar5G:18} and/or the coding rate~\cite{Andrews:2011bg,JHParkTWC:15}. In addition, we assume separate coding for each miniblock, which may loose frequency diversity gain compared to the coding across multiple miniblocks \cite{Petar5G:18,TseBook:FundamaentalsWC:2005}. 

With these assumptions and the $\SIR$s that are identically distributed across miniblocks, average rate is determined by the decoding success probability for any single miniblock. Therefore, we drop the superscript $(k)$ in $\SIR_{i,j}^{(k)}$ and the small-scale fading terms, and derive the average rate in the sequel.

\subsubsection{OMA}
The typical user can decode the signal from $\L_i$ with the decoding success probability $p_{i,\OMA}(t_{i,\OMA})$ that equals

\vspace{-10pt}\small\begin{align}
p_{i,\OMA} &= \Pr\big(\log(1+\SIR_{i,\OMA})\geq r_{i,\OMA}  \big)\\
&= \Pr\(\SIR_{i,\OMA} \geq t_{i,\OMA}\), \label{Eq:success_OMA}
\end{align}\normalsize
where $r_{i,\OMA}$ is the coding rate per miniblock, which is hereafter rephrased as a target $\SIR$ threshold $t_{i,\OMA}= e^{r_{i,\OMA}}-1$.

For the given target decoding success probability $\eta_i$ of $\L_i$, the average rate $R_{i,\OMA}(\eta_i)$ of $\L_i$ is obtained by using outage capacity~\cite{TseOC:07} as

\vspace{-10pt}\small\begin{align}
R_{i,\OMA}(\eta_i) &= w_{i,\OMA} \eta_i  \cdot \sup\{\log(1 + t_{i,\OMA}): p_{i,\OMA}(t_{i,\OMA})\geq \eta_i\} \\
		&= w_{i,\OMA} \eta_i \cdot \log(1 + t_{i,\OMA}^* ), \label{Eq:rate_OMA}
\end{align}\normalsize
where the optimal target $\SIR$ threshold $t_{i,\OMA}^*$ satisfies $p_{i,\OMA}(t_{i,\OMA}^*)=\eta_i$, and thus equals $t_{i,\OMA}^* = p_{i,\OMA}^{-1}(\eta_i)$. 

Note that even when the coding block length of $\L_1$ is short, the average rate expression in~\eqref{Eq:rate_OMA} still holds, since the finite-block length rate under fading channels converges to the outage capacity~\cite{DurisiPolyanski:14}.

\subsubsection{NOMA}
With the SIC order that decodes $\L_1$ prior to $\L_2$, the typical user's decoding success probabilities $p_{1,\NOMA}(t_{1,\NOMA})$ and $p_{2,\NOMA}(t_{1,\NOMA},t_{2,\NOMA})$ of $\L_1$ and $\L_2$ are given~as

\vspace{-10pt}\small\begin{align}
p_{1,\NOMA}(t_{1,\NOMA}) &= \Pr\(\SIR_{1,\NOMA}\geq t_{1,\NOMA}\) \quad\text{and} \label{Eq:success1_NOMA}\\
p_{2,\NOMA}(t_{1,\NOMA},t_{2,\NOMA}) &= p_{1,\NOMA}(t_{1,\NOMA})\cdot \Pr\(\SIR_{2,\NOMA}\geq t_{2,\NOMA}\mid \SIR_{1,\NOMA}\geq t_{1,\NOMA}\). \label{Eq:success2_NOMA}
\end{align}\normalsize
Following \cite{JindalSIC:09}, our SIC do not allow to decode the $\L_2$ signal after the decoding failure of the $\L_1$ signal. With a different SIC architecture that allows such a decoding attempt,~\eqref{Eq:success2_NOMA} is regarded as the lower bound, as done in~\cite{Petar5G:18}.

For the given target decoding success probability $\eta_1$ of $\L_1$, the average rate $R_{1,\NOMA}(\eta_1)$ of $\L_1$ is given as

\vspace{-10pt}\small\begin{align}
R_{1,\NOMA}(\eta_1) &= \eta_1 \cdot  \sup\{\log(1 + t_{1,\NOMA}): p_{1,\NOMA}(t_{1,\NOMA})\geq \eta_1\} \label{Eq:rate_NOMA_pre}\\
&= \eta_1 \cdot  \log(1 + t_{1,\NOMA}^*), \label{Eq:rate_NOMA}
\end{align}\normalsize
where the optimal target $\SIR$ threshold equals $t_{1,\NOMA}^* = p_{1,\NOMA}^{-1}(\eta_1)$. Similarly, for the given target decoding success probability $\eta_2$ of $\L_2$, the average rate $R_{2,\NOMA}(\eta_1,\eta_2)$ of $\L_2$ is given as

\vspace{-10pt}\small\begin{align}
\hspace{-5pt} R_{2,\NOMA}(\eta_1,\eta_2) &= \eta_2 \cdot  \sup\{\log(1 + t_{2,\NOMA}): p_{2,\NOMA}(t_{1,\NOMA}^*, t_{2,\NOMA})\geq \eta_2\} \label{Eq:rate_NOMA_pre}\\
&= \eta_2 \cdot  \log(1 + t_{2,\NOMA}^*), \label{Eq:rate_NOMA2}
\end{align}\normalsize
where $t_{2,\NOMA}^*=p_{2,\NOMA}^{-1}(\eta_1, \eta_2)$.

\subsection{Visuo-Haptic Perceptual Resolution} \label{Sect:Sys_JND}
The resolution of human perceptions is often measured by using JND in psychophysics. In a psychophysical experiment, the JND is calculated as the minimum stimulus variation that can be detectable during $84$\% of the trials~\cite{Ernst:2002aa}. For a visuo-haptic perception, its integrated JND is obtained by combining the JNDs of visual and haptic perceptions. 

To elaborate, when individual haptic and visual perceptions have the perceived noise variances $\sigma_1$ and $\sigma_2$, a human brain combines these perceptions, yielding an integrated noise variance $\sigma_{12}$ that satisfies $\sigma_{12}^{-2} = \sigma_1^{-2} + \sigma_2^{-2}$. This relationship was first discovered in~\cite{Ernst:2002aa} by measuring the corresponding JNDs that are proportional to the perceived noise variances. The said relationship is thus read as $\gamma_{12}^{-2}=\gamma_1^{-2} + \gamma_2^{-2}$, where $\gamma_{12}$ denotes the JND measured when using both visuo-hapric perceptions, while $\gamma_{1}$ and $\gamma_2$ identify the JNDs of the individual haptic and visual perceptions, respectively.

For individual visual perceptions, it has been reported by another experiment~\cite{Shi:10} that the PER is proportional to its sole JND $\gamma_2$ due to the resulting visual frame loss. Similarly, for individual haptic perceptions, it has been observed in~\cite{ShiHirche:16} that the PER is proportional to the elapsed time to complete a given experimental task, which increases with the corresponding JND $\gamma_1$ due to the coarse perceptions. Based on such experimental evidence, we can write that $\gamma_i=(1-\eta_i)^2$, where $(1-\eta_i)$ represents the PER on $\L_i$. 

Accordingly, the JND $\gamma_{12}$ of visuo-haptic perceptions is obtained from the following equation

\vspace{-10pt}\small\begin{align}
\gamma_{12}^{-2} = (1-\eta_1)^{-2} + (1-\eta_2)^{-2}. \label{Eq:JND}
\end{align}\normalsize
In the following subsection, we adjust the target decoding success probabilities $\eta_1$ and $\eta_2$ of $\L_1$ and $\L_2$, so as to guarantee a target visuo-haptic JND $\theta>0$, i.e., $\gamma_{12}=\theta$.

\subsection{URLLC-eMBB Multiplexing Problem Formulation} \label{Sect:Sys_Problem}
In a downlink cellular system serving visuo-haptic VR traffic, haptic and visual perceptions are supported through $\L_1$ and $\L_2$, respectively. Each link pursues different service objectives as follows. The URLLC $\L_1$ aims at:
\begin{enumerate}[(i)]
\item Ensuring a target decoding success probability $\eta_1<1$; and
\item Ensuring a target average rate $\R > 0$.
\end{enumerate}
In contrast, the eMBB $\L_2$ aims at:
\begin{enumerate}[(i)]\addtocounter{enumi}{2}
\item Maximizing the average rate $R_{2,j}(\eta_i) > 0$; while
\item Ensuring a target decoding success probability $\eta_2$, with $\hat{\eta}_2 \leq \eta_2<\eta_1$.
\end{enumerate}
In (iv), $\hat{\eta}_2\leq \eta_2$ follows from an experimental evidence that the quality of visual perceptions dramatically drops when PER~exceeds a certain limit, e.g., $10$\% PER that equals~$\hat{\eta}_2 = 0.9$~\cite{Shi:10}.

In addition to these individual service objectives, with $\L_1$ and $\L_2$, their aggregate JND $\gamma_{12}$ should guarantee a target visuo-haptic JND $\theta$. The said service objectives and requirements of $\L_1$ and $\L_2$ are described in the following problem formulation.

\vspace{-10pt}\small\begin{subequations}
\begin{align}
\hspace{-1.3cm}\textsf{(P1)} \;\;\;\; \underset{\{W_{i,j},\; \beta_{i,j},\; t_{i,j},\; \eta_i\} }{\textsf{maximize}}&\;\; R_{2,j}(\eta_1,\eta_2)\\[-1mm]
&\hspace{-25pt}\textsf{s.t.}\;\;\;\;\; R_{1,j}(\eta_1) = \R  \label{Eq:Const1}\\
&\hspace{-9pt}\;\;\;\; \gamma_{12} = \theta \label{Eq:Const3}\\
&\hspace{-5pt}\;\;\;\; \hat{\eta}_2\leq \eta_2 < \eta_1 < 1 \label{Eq:Const4}
\end{align}
\end{subequations}\normalsize
The objective functions $R_{2,j}(\eta_1,\eta_2)$ and $R_{1,j}(\eta_1)$ in the constraint \eqref{Eq:Const1} are obtained from \eqref{Eq:rate_OMA} for OMA and from \eqref{Eq:rate_NOMA} for NOMA. In the constraint~\eqref{Eq:Const3}, $\gamma_{12}$ is provided in~\eqref{Eq:JND}. Without loss of generality, we hereafter consider a sufficiently large number $K$ of miniblocks so that the miniblock allocation fraction $w_{i,j}$ under OMA is treated as a continuous value.

\section{Optimal Multiplexing of Visuo-Haptic VR Traffic under OMA and NOMA}
In this section, we optimize the multiplexing of $\L_1$ and $\L_2$ that support visuo-haptic VR traffic.
With \textsf{P1}, for OMA, we optimize the miniblock allocation from the unit frequency block to each link. For NOMA, on the other hand, we optimize the power allocation from the unit transmit power.

\begin{figure*}
\centering
\subfigure[With $\theta=10^{-4}$.]{\includegraphics[width= \figwidth]{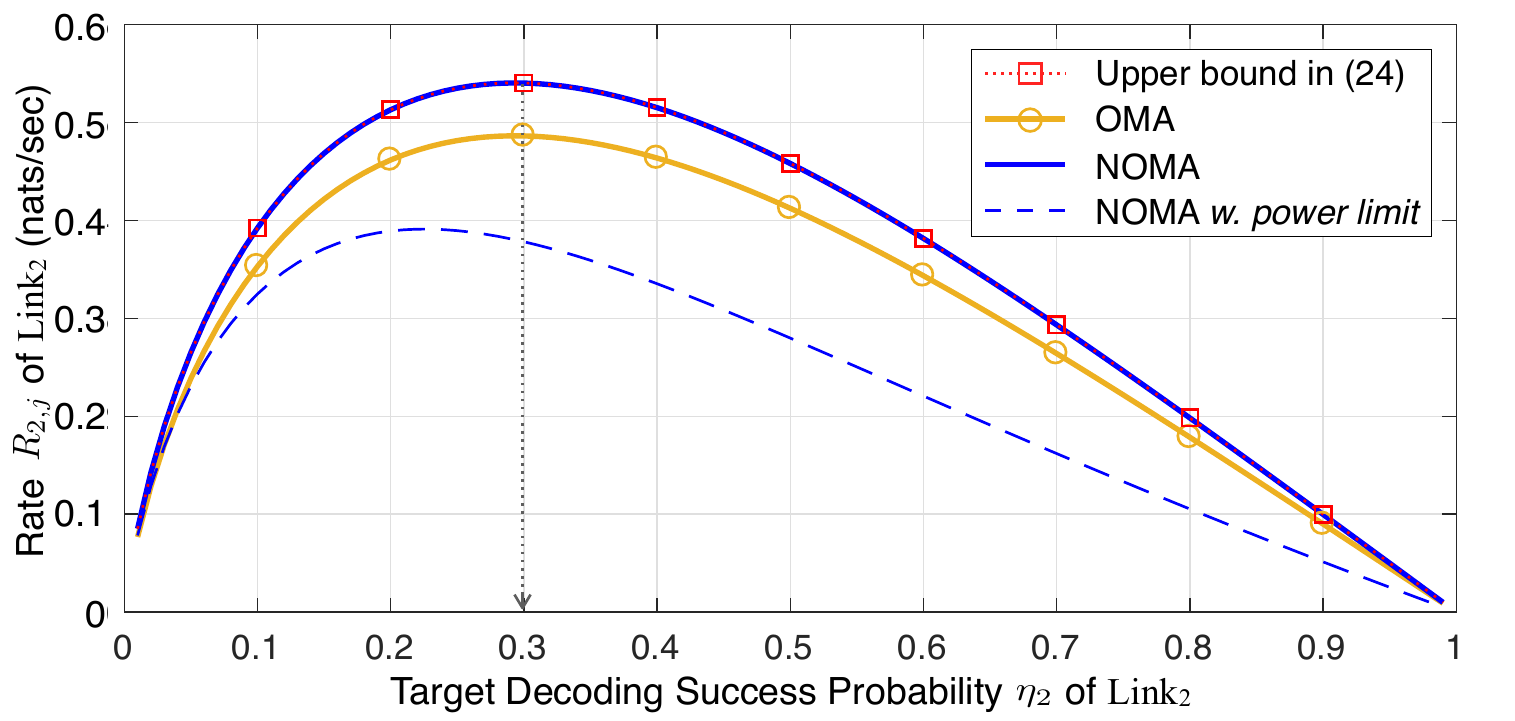}}\hspace{15pt}
\subfigure[With $\theta=10^{-2}$.]{\includegraphics[width=  \figwidth]{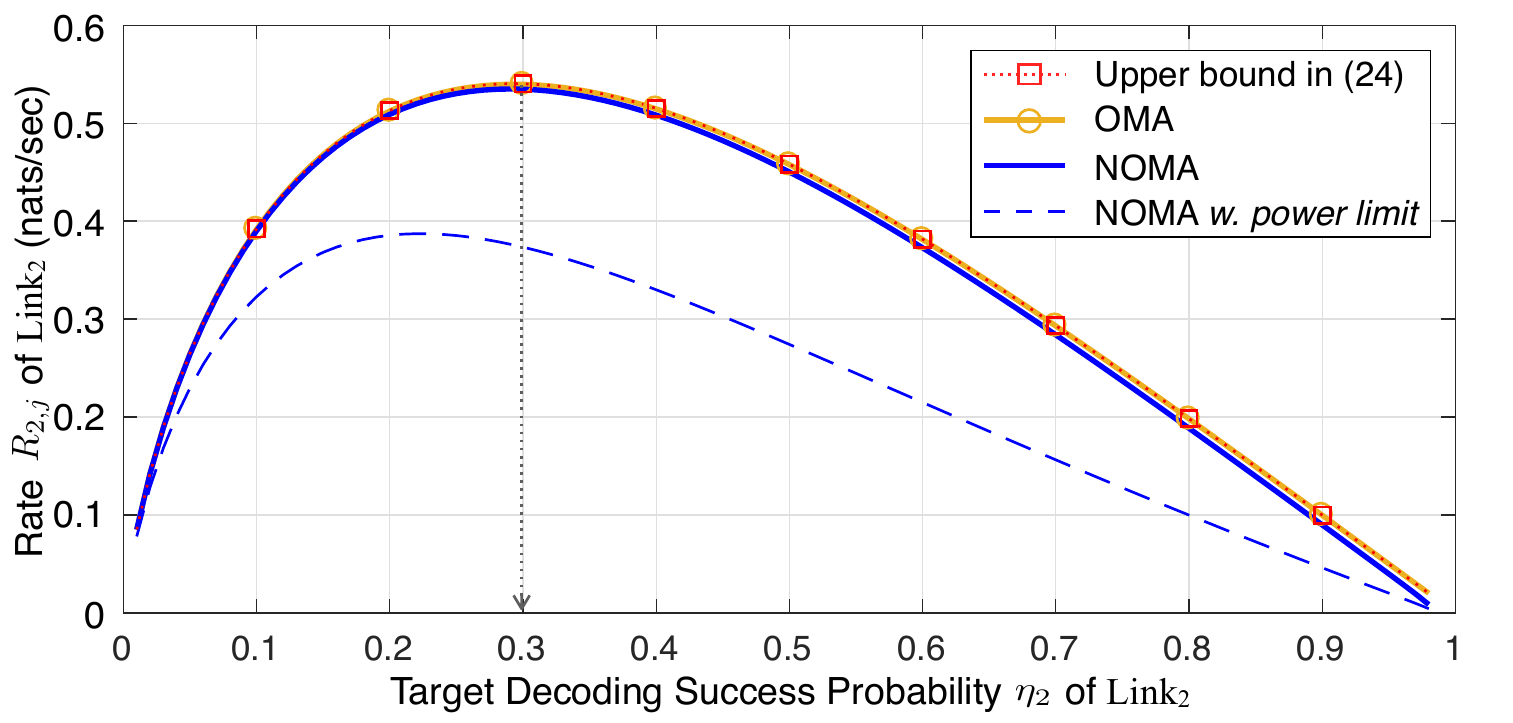}}
\caption{\small Average rate $R_{2,j}^*(\eta_2)$ w.r.t. the target decoding success probability $\eta_2$ under OMA and NOMA ($\R=10^{-5}$ nats/sec, $\alpha=4$).}
\end{figure*}

\subsection{Optimal OMA} \label{Sect:OptOMA}
We aim at optimizing the miniblock allocation $w_{i,\OMA} < 1$. To this end, for given $w_{i,\OMA}$ and $\eta_i$, we derive the average rate $R_{i,\OMA}(\eta_i)=w_{i,\OMA}\eta_i  \log(1 + t_{i,\OMA}^*)$ with $t_{i,\OMA}^*=p^{-1}_{i,\OMA}(\eta_i)$. This requires taking the inverse function of $p_{i,\OMA}(t_{i,\OMA})$ in \eqref{Eq:success_OMA}. 

The typical user's $p_{i,\OMA}(t_{i,\OMA})=\Pr(\SIR_{i,\OMA}\geq t_{i,\OMA})$ is commonly referred to as $\SIR$ coverage probability, and its closed-form expression can be derived by using stochastic geometry \cite{Andrews:2011bg,Haenggi:ISIT14}. Namely, $p_{i,\OMA}(t_{i,\OMA})$ is given as

\footnotesize\vspace{-10pt}\begin{align}
p_{i,\OMA}(t_{i,\OMA}) &= \E_{x_o,\Phi_o}\[\Pr\(g_i>  t_{i,\OMA} |x_o|^{\alpha} \sum_{x\in\Phi_o } g_{x} |x|^{-\alpha}\)\]\\
&= \E_{x_o,\Phi_o}\[\prod_{x\in\Phi_o }\exp\(-  t_{i,\OMA}  g_{x} \(\frac{|x_o|}{|x|}\)^\alpha \) \] \label{Eq:SG_OMA1}\\
&= \E_{x_o,\Phi_o}\[\prod_{x\in\Phi_o } \(1 + t_{i,\OMA} \(\frac{|x_o|}{|x|}\)^\alpha \)^{-1} \], \label{Eq:SG_OMA2}
\end{align}\normalsize
where \eqref{Eq:SG_OMA1} comes from the complementary cumulative density function (CCDF) of $g$, and \eqref{Eq:SG_OMA2} follows from the Laplace functional of the i.i.d. exponential variables $g_x$'s. Applying the probability generating functional (PGFL) of a stationary PPP $\Phi_o$ \cite{HaenggiSG}, we obtain the integral expression

\footnotesize\begin{align}
\hspace{-10pt}\eqref{Eq:SG_OMA2} &= \E_{x_o}\[ \exp\( -2 \pi \lambda\int_{v>|x_o|}\(1 + \frac{1}{t_{i,\OMA}}\(\frac{v}{|x_o|}\)^\alpha \)v dv  \)\]\\ 
&= 1/{}_2F_1\(1, -2/\alpha; 1-2/\alpha; -t_{i,\OMA}\). \label{Eq:SG_hypergeo}
\end{align}\normalsize
The last step is derived by using the void probability of $\Phi$, with the definition of a Gauss hypergeometric function that equals ${}_2F_1(a, b;c;z)=\sum_{n=0}^\infty \frac{\Gamma(a + n)\Gamma(b+n)\Gamma(c)}{\Gamma(a)\Gamma(b)\Gamma(c+n)} \frac{z^{n}}{n!}$ where $\Gamma(x)$ is the gamma function.

In spite of the closed-form $\SIR$ coverage probability expression in \eqref{Eq:SG_hypergeo}, due to the hypergeometric function, the inverse function of $p_{i,\OMA}(t_{o,\OMA})$ can only be numerically computed. We resolve this problem by exploiting a simplified $\SIR$ coverage probability bound, proposed in our previous study~\cite{RelCovWCL:17}.
\begin{lemma} \linespread{\thmspacing} (Closed-form $\SIR$ coverage bounds) \emph{Denoting as $\SIRn= g|x_o|^{-\alpha}/\sum_{x\in\Phi_o}g_x |x|^{-\alpha}$, according to Theorem~1 and Corollary~1~in~\cite{RelCovWCL:17}, the coverage probability of $\SIRn$ is upper and lower bounded as
\small\begin{align}
\Pr(\SIRn\geq t) = (1 + c t)^{-\frac{2}{\alpha}},
\end{align}\normalsize
where ${\alpha}/{(\alpha-2)} \leq c \leq \[{2\pi}/{\alpha}\cdot \csc\({2\pi}/{\alpha}\)\]^{{\alpha}/{2}}$.
}\end{lemma}
\noindent  As validated in~\cite{RelCovWCL:17}, these upper and lower bounds guarantee the convergence to the exact values respectively for $t\rightarrow 0$ and $t\rightarrow \infty$, which corresponds to high and low target decoding success probabilities, respectively. We henceforth treat the bounds as the approximated coverage probability of $\SIRn$.

Applying this to $p_{i,\OMA}(t_{i,\OMA})=\Pr(\SIRn \geq t_{i,\OMA})$, we obtain $p_{i,\OMA}(t_{i,\OMA})=(1+c  t_{i,\OMA})^{-2/\alpha}$. We thereby derive the inverse function $t_{i,\OMA}^*=p_{i,\OMA}^{-1}(\eta_i)$ that equals

\vspace{-10pt}\small\begin{align}
t_{i,\OMA}^* = G(\eta_i), \label{Eq:Optt}
\end{align}\normalsize
where $G(\eta_i)=({\eta_i}^{-\alpha/2}-1)/c$. The specific value of $c$ and the approximation accuracy are to be specified in Sect.~IV.

Applying \eqref{Eq:Optt} to $R_{i,\OMA}(\eta_i)$ in \eqref{Eq:rate_OMA} yields the following result.
\begin{proposition}\linespread{\thmspacing} (Closed-form average rate, OMA) \emph{For a given~$\eta_i$, the average rate of $\L_i$ under OMA is given as
\small\begin{align}
R_{i,\OMA}(\eta_i) &= w_{i,\OMA} \eta_i  \log\(1 + {G(\eta_i)} \).
\end{align}\normalsize
}\end{proposition}

Finally, applying this result to \textsf{P1}, we obtain the optimal OMA design and its corresponding average rate as below.
\begin{proposition} 
\linespread{\thmspacing}
(Maximum average rate, OMA) \emph{The maximum average rate of $\L_2$ under OMA is provided as
\small\begin{align}
R_{2,\OMA}^*(\eta_2^*) &= w_{2,\OMA}^* \eta_2^*  \log\(1 + {G(\eta_2^*)} \) \quad\quad \text{where} \label{Eq:OptRate_OMA}\\
w_{2,\OMA}^* &= 1-\frac{\R}{\eta_1 \log(1 + G(\eta_1^*))}, \label{Eq:OptW2}\\
\eta_1^* &= 1- \(\frac{1}{\theta^2}-\frac{1}{(1-\eta_2^*)^2}\)^{-\frac{1}{2}},\\
\eta_2^* &= \underset{\hat{\eta}_2 \leq \eta_2 < \min(1,1-\theta \sqrt{2})}{\arg\max}  R_{2,\OMA}^*(\eta_2), \label{Eq:Prop2Opteta2}
\end{align}\normalsize
$R_{2,\OMA}^*(\eta_2) = w_{2,\OMA}^*(\eta_2)\eta_2 \log\(1 + {G(\eta_2)}\)$, and $w_{2,\OMA}^*(\eta_2)$ is obtained by replacing $\eta_2^*$ with $\eta_2$ in $w_{2,\OMA}^*$.\\
\begin{proof} See Appendix-A.
\end{proof}
}\end{proposition}
\noindent Notice that $R_{2,\OMA}^*(\eta_2^*)$ still requires optimization with respect to $\eta_2$, as shown in \eqref{Eq:Prop2Opteta2}. In fact, the objective function to be maximized in \eqref{Eq:Prop2Opteta2} is concave over $\eta_2$, as $w_{2,\OMA}^*(\eta_2)$ is a monotone increasing function of~$\eta_2$. Nevertheless, it is non-differentiable due to $w_{2,\OMA}^*(\eta_2)$, and can thus be optimized only by simulation.

In order to obtain the closed-form expression of $R_{2,\OMA}^*(\eta_2^*)$, we consider a sufficiently large $\hat{\eta}_2$ in~\eqref{Eq:Const4}, such that the concave objective function~in \eqref{Eq:Prop2Opteta2} monotonically decreases with $\eta_2\geq \hat{\eta}_2$. In this regime, the optimum $R_{2,\OMA}^*(\eta_2^*)$ is achieved at~$\eta_2^*=\hat{\eta}_2$, yielding the following corollary.
\begin{corollary}
\linespread{\thmspacing}
(Closed-form maximum average rate, OMA) \emph{For $\hat{\eta}_2\rightarrow 1$, the objective function in \eqref{Eq:Prop2Opteta2} is a monotone decreasing function of $\eta_2$, and we thus obtain $R_{2,\OMA}^*(\eta_2^*)=R_{2,\OMA}^*(\hat{\eta}_2)$.
}\end{corollary}
\noindent Note that Corollary 1 holds even under $\hat{\eta}_2$ that is much smaller than $1$. For instance, as illustrated by simulation in Fig.~2, $R_{2,\OMA}^*(\eta_2)$ monotonically decreases with $\eta_2\geq 0.3$. This value is still much smaller than a practical value of $\hat{\eta}_2$, e.g., $\hat{\eta}_2=0.9$ as reported in \cite{Shi:10,UR2Cspaswin:17}. Therefore, we conclude that $R_{2,\OMA}^*(\hat{\eta}_2)$ in Corollary~1 is the closed-form approximation. Validating its accuracy compared to Proposition~1 is deferred to Sect.~IV.

\begin{figure*}%
\centering
\subfigure[With $\theta=10^{-4}$.]{
\includegraphics[width=  \figwidth ]{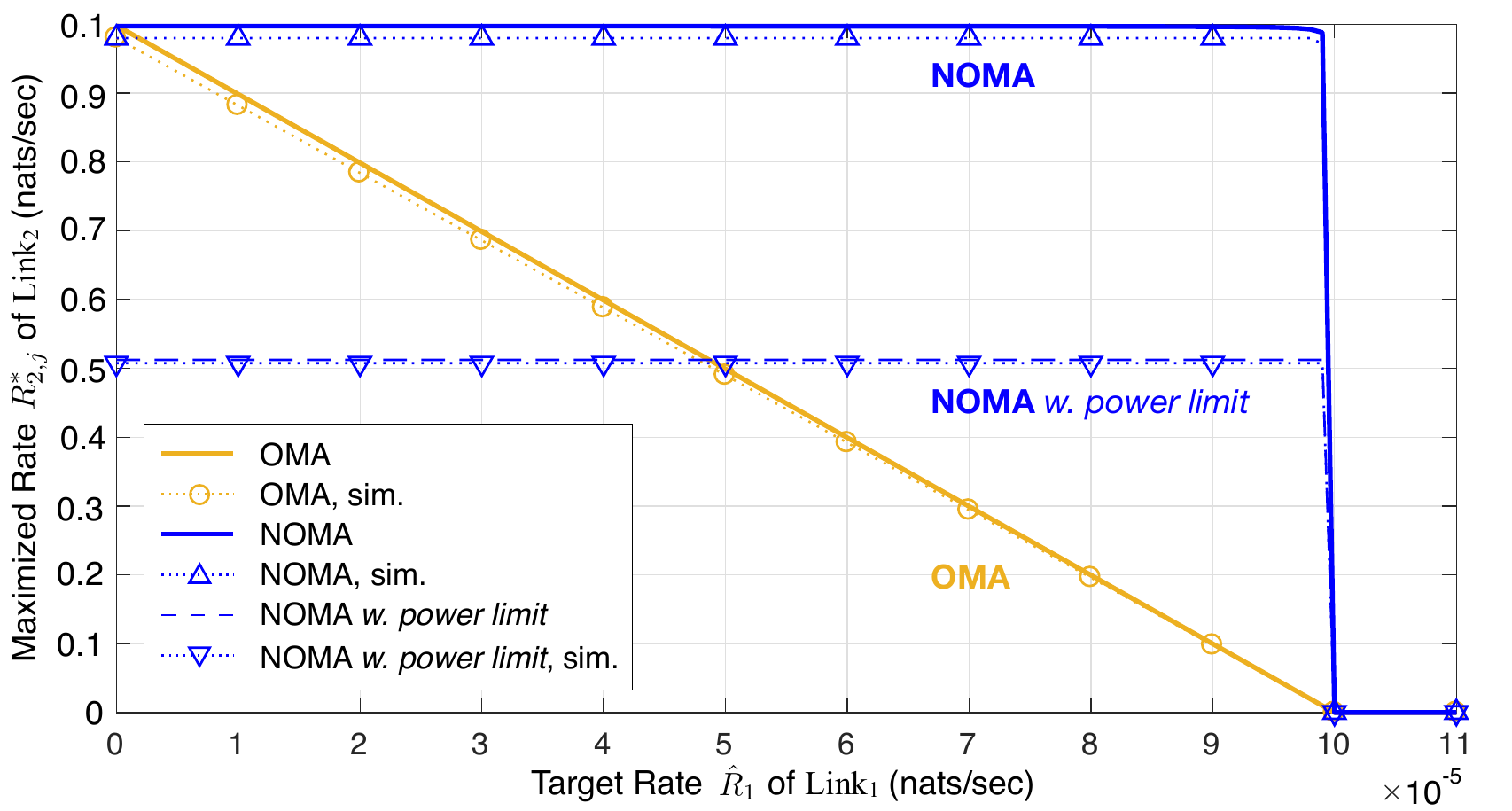}} \hspace{15pt}
\subfigure[With $\theta=10^{-2}$.]{
\includegraphics[width=  \figwidth ]{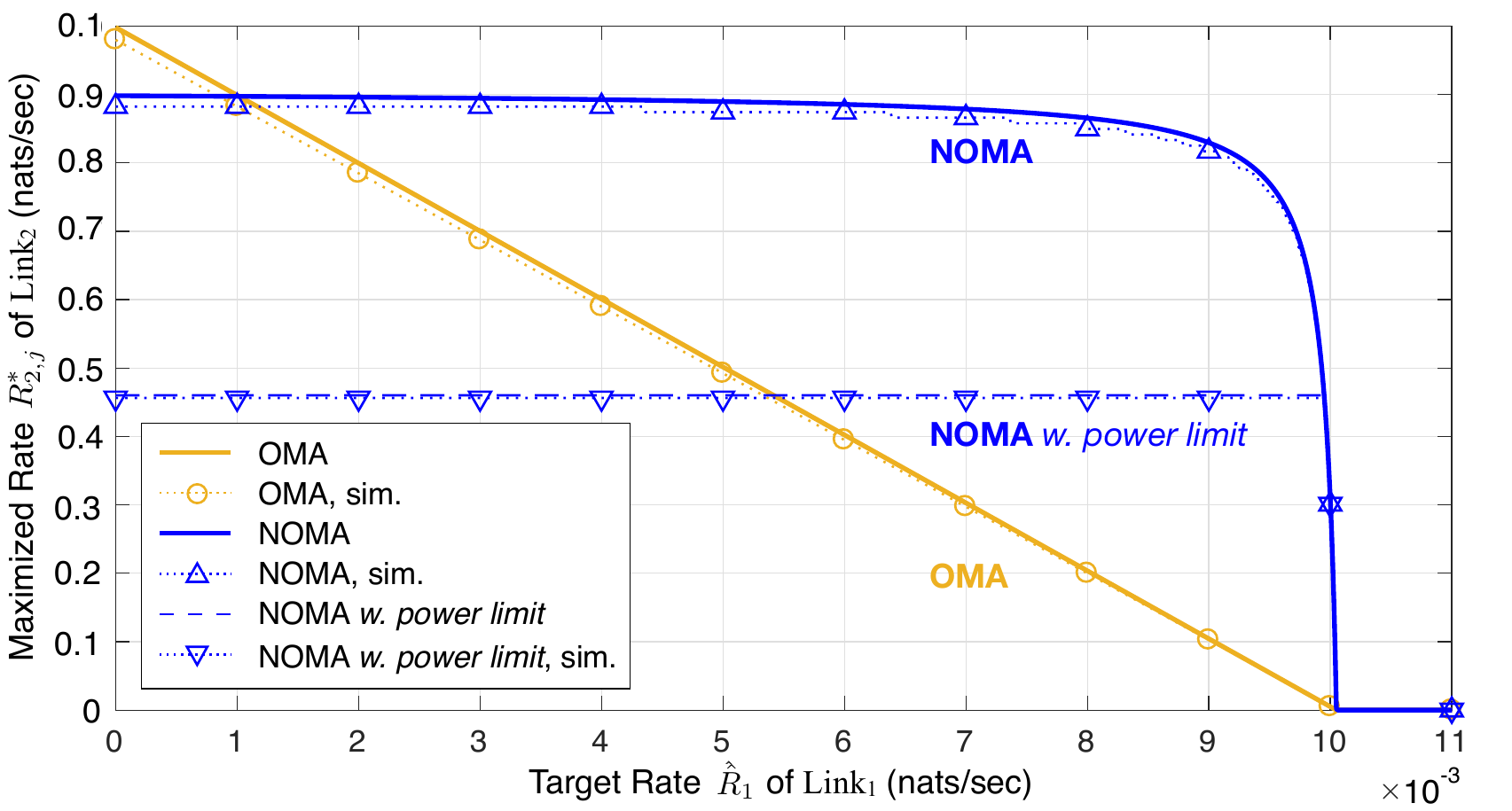}}%
\caption{\small{Maximized average rates $R_{2,\OMA}^*(\eta_2^*)$ and $R_{2,\NOMA}^*(\eta_1^*,\eta_2^*)$ w.r.t. a target rate $\R$ of $\L_1$ for different target visuo-haptic JND~thresholds.}}
\end{figure*}

\subsection{Optimal NOMA} \label{Sect:OptNOMA}
Our goal is to optimize the transmit power allocation fraction $\beta_{i,\NOMA}\leq 1$ to $\L_i$. To this end, for given $\beta_{i,\NOMA}$ and $\eta_i$, we derive the average rate $R_{i,\NOMA}(\eta_i)$ provided in \eqref{Eq:rate_NOMA}. The average rate is determined by the decoding order of the superposition of $\L_1$ and $\L_2$ signals. We propose the following optimal SIC decoding order.
\begin{lemma}
\linespread{\thmspacing}
(Reliability-ordered SIC) \emph{In \textsf{(P1)} with $\eta_1>\eta_2$, $R_{2,\NOMA}(\eta_2)$ is maximized when \emph{$\L_1$ is decoded prior to $\L_2$}.\\
\begin{proof} 
By contradiction, we assume $\L_2$ is decoded prior to $\L_1$. We obtain the corresponding $p_{1,\NOMA}(t_{1,\NOMA},t_{2,\NOMA})$ by exchanging the subscripts $1$ and~$2$ in \eqref{Eq:success2_NOMA}, given as
\small\begin{align}
p_{1,\NOMA}(t_{1,\NOMA},t_{2,\NOMA}) = p_{2,\NOMA}(t_{2,\NOMA})\cdot \Pr(\SIR_{1,\NOMA}\geq t_{1,\NOMA}\mid \SIR_{2,\NOMA}\geq t_{2,\NOMA}).
\end{align}\normalsize\\[-5mm]
According to \eqref{Eq:rate_NOMA_pre}, it should guarantee $p_{1,\NOMA}(t_{1,\NOMA},t_{2,\NOMA})\geq \eta_1$, which now reads\
\small\begin{align}
\Pr(\SIR_{1,\NOMA}\geq t_{1,\NOMA}\mid \SIR_{2,\NOMA}\geq t_{2,\NOMA}) \geq {\eta_1}/{p_{2,\NOMA}(t_{2,\NOMA})}. \label{Eq:Lemma2pf}
\end{align}\normalsize
Since $\Pr(\SIR_{1,\NOMA}\geq t_{1,\NOMA}\mid \SIR_{2,\NOMA}\geq t_{2,\NOMA})\leq 1$, the denominator of the RHS in \eqref{Eq:Lemma2pf} should satisfy $p_{2,\NOMA}(t_{2,\NOMA})\geq \eta_1$. On $\L_2$, this consequently imposes the target decoding success probability at least $\eta_1$, which is higher than its original target~$\eta_2$. The resulting $R_{2,\NOMA}(\eta_1)$ is thus smaller than $R_{2,\NOMA}(\eta_2)$ in \eqref{Eq:rate_NOMA} with the opposite decoding order, finalizing the~proof.
\end{proof}
}\end{lemma}

The proposed downlink SIC in order of reliability is consistent with the same SIC rule for uplink URLLC-eMBB multiplexing, proposed in \cite{Petar5G:18}. Note that this reliability-ordered SIC may not always comply with a traditional SIC design that decodes the stronger signal first~\cite{3GPPMUST:2015,TseBook:FundamaentalsWC:2005}. If $\beta_{1,\NOMA}>\beta_{2,\NOMA}$, i.e., $\beta_{1,\NOMA}>0.5$, then the reliability-ordered SIC follows the traditional power-ordered SIC. Otherwise, the reliability-ordered SIC should decode the weaker signal first, which is not allowed under the power-ordered SIC. Such restriction is examined by simulation in Sect.~IV.

With the reliability-ordered SIC, as done for OMA in Sect.~\ref{Sect:OptOMA}, we exploit Lemma~1, and derive the closed-form decoding success probabilities, followed by the average rates.

First, when decoding the signal intended for $\L_1$, the decoding success probability $p_{1,\NOMA}(t_{1,\NOMA})=\Pr\(\SIR_{1,\NOMA} \geq t_{1,\NOMA}\)$ is rephrased as $p_{1,\NOMA}(t_{1,\NOMA})=\Pr(\SIRn \geq [1/t_{i,\NOMA}-\beta_{2,\NOMA}/\beta_{1,\NOMA}]^{-1})$, which follows from \eqref{Eq:success1_NOMA} with $\SIR_{1,\NOMA}$ in \eqref{Eq:SIR1_NOMA}. Applying Lemma~1 to this, we thereby obtain

\vspace{-10pt}\small\begin{align}
\hspace{-5pt} p_{1,\NOMA}(t_{1,\NOMA}) &=  1-\Big(1 + c\Big[{1}/{t_{1,\NOMA}}-{\beta_{2,\NOMA}}/{\beta_{1,\NOMA}}\Big]^{-1}\Big)^{-{2}/{\alpha}}. \label{Eq:Prop3pf1}
\end{align}\normalsize
By taking the inverse function, $t_{1,\NOMA}^*=p_{1,\NOMA}^{-1}(\eta_1)$ is derived as

\vspace{-10pt}\small\begin{align}
t_{1,\NOMA}^* = \Big[\beta_{2,\NOMA}/\beta_{1,\NOMA} + 1/G(\eta_1) \Big]^{-1}. \label{Eq:t1NOMAOpt}
\end{align}\normalsize
Applying this to $R_{1,\NOMA}(\eta_1)$ in \eqref{Eq:rate_NOMA}, we finally obtain the closed-form average rate, provided in Proposition~3 on the next page.

Next, when decoding the signal intended for $\L_2$, following from \eqref{Eq:success2_NOMA} with $\SIR_{2,\NOMA}$ in \eqref{Eq:SIR2_NOMA} and from $p_{1,\NOMA}(t_{1,\NOMA}^*)=\eta_1$ in~\eqref{Eq:rate_NOMA}, the decoding success probability $p_{2,\NOMA}(t_{1,\NOMA}^*, t_{2,\NOMA})$ is

\vspace{-10pt}\small\begin{align}
&\hspace{-5pt} p_{2,\NOMA}(t_{1,\NOMA}^*,t_{2,\NOMA}) = \eta_1 \Pr\(\SIR_{2,\NOMA} \geq t_{2,\NOMA}\mid\SIR_{1,\NOMA} \geq t_{1,\NOMA}^*\) \label{Eq:NOMA_p2first}\\
\hspace{5pt}&= \eta_1 \Pr\(\SIRn \geq \max\{t_{2,\NOMA}/\beta_{2,\NOMA}, [1/t_{1,\NOMA}^* - \beta_{2,\NOMA}/\beta_{1,\NOMA} ]^{-1} \} \) \label{Eq:NOMA_p2cond}\\
\hspace{5pt}&\approx \eta_1 \Pr\(\SIRn \geq t_{2,\NOMA}/\beta_{2,\NOMA} \) = \eta_1 \Pr(\SIR_{2,\NOMA}\geq t_{2,\NOMA}). \label{Eq:NOMA_p2final}
\end{align}\normalsize
In \eqref{Eq:NOMA_p2cond}, we utilize the relationship $\Pr(\SIR_{2,\NOMA}\geq t_{2,\NOMA})=\Pr(\SIRn \geq t_{2,\NOMA}/\beta_{2,\NOMA})$, as well as $\Pr\(\SIR_{1,\NOMA}\geq t_{1,N}^*\)=\Pr(\SIRn \geq [1/t_{i,\NOMA}^*-\beta_{2,\NOMA}/\beta_{1,\NOMA}]^{-1})$. The approximation in \eqref{Eq:NOMA_p2final} is valid for high $\eta_{1}$, since $t_{1,\NOMA}^*$ in \eqref{Eq:t1NOMAOpt} approaches $0$ as $\eta_1\rightarrow 1$. 

By comparing \eqref{Eq:NOMA_p2final} and \eqref{Eq:NOMA_p2first}, we conclude that $\L_2$ signal is almost independently decoded with $\L_1$ decoding under high $\eta_1$. In fact, $\SIR_{1,\NOMA}$ in \eqref{Eq:SIR1_NOMA} and $\SIR_{2,\NOMA}$ in \eqref{Eq:SIR2_NOMA} are correlated over both large-scale fading and small-scale fading, and the $\L_1$ decoding success may thus affect the subsequent $\L_2$ decoding success. With high $\eta_1$, nevertheless, $\L_1$ can be decoded almost regardless of the channel quality, negligibly contributing to the $\L_2$ decoding success.

Finally, applying Lemma~1 to \eqref{Eq:NOMA_p2final}, we obtain $t_{2,\NOMA}^*=p_{2,\NOMA}^{-1}(\eta_1,\eta_2)$ that equals 

\vspace{-10pt}\small\begin{align}
t_{2,\NOMA}^* = \beta_{2,\NOMA} H(\eta_1,\eta_2),
\end{align}\normalsize
where $H(\eta_1,\eta_2)=[(\eta_1/\eta_2)^{\alpha/2}-1]/c$. Plugging this in \eqref{Eq:rate_NOMA2} yields the closed-form average rate as follows.
\begin{proposition}
\linespread{\thmspacing}
(Closed-form average rate, NOMA) \emph{For a given $\eta_i$, the average rate of $\L_i$ under NOMA with the reliability-ordered SIC is given~as
\small\begin{align}
R_{1,\NOMA}(\eta_1) &= \eta_1 \log\(1 + \[\frac{\beta_{2,\NOMA}}{\beta_{1,\NOMA}} + \frac{1}{G(\eta_1)}\]^{-1} \) \quad\text{and} \label{Eq:Prop3Rate1_NOMA}\\
R_{2,\NOMA}(\eta_1,\eta_2) &= \eta_2  \log\Big(1 + \beta_{2,\NOMA}H(\eta_1,\eta_2)\Big). \label{Eq:Prop3Rate2_NOMA}
\end{align}\normalsize
}\end{proposition}

With these closed-form average rate expressions, we solve \textsf{P1}, and obtain the optimal NOMA design as follows.
\begin{proposition}
\linespread{\thmspacing}
(Maximum average rate, NOMA) \emph{The maximum average rate of $\L_2$ under NOMA is provided as
\small\begin{align}
\hspace{-10pt} R_{2,\NOMA}^*(\eta_1^*,\eta_2^*) =& \eta_2  \log\Big(1 + \beta_{2,\NOMA}^*H(\eta_1^*,\eta_2^*)\Big)  \quad \text{where} \label{Eq:OptRate_NOMA}\\
\hspace{-10pt} \beta_{2,\NOMA}^* =& \(1 + \[\frac{1}{\exp\({\R }/{ \eta_1^* }\) -1} -\frac{1}{G(\eta_1^*)}\] ^{-1}  \)^{-1}\hspace{-10pt},\\ \label{Eq:Optbeta2}
\hspace{-10pt} \eta_1^* =& 1- \(\frac{1}{\theta^2}-\frac{1}{(1-\eta_2^*)^2}\)^{-\frac{1}{2}},\\
\hspace{-10pt}\eta_2^* =& \underset{\hat{\eta}_2 \leq \eta_2 < \min(1,1-\theta \sqrt{2})}{\arg\max}  R_{2,\NOMA}^*(\eta_1,\eta_2), \label{Eq:Prop4last}
\end{align}\normalsize
$R_{2,\NOMA}^*(\eta_1,\eta_2)=\eta_2 \log\( 1 + \beta_{2,\NOMA}^*(\eta_2)H(\eta_1,\eta_2) \)$, and $\beta_{2,\NOMA}^*(\eta_2)$ is obtained by replacing $\eta_2^*$ in $\beta_{2\NOMA}^*$ with $\eta_2$.
\\\begin{proof} See Appendix-B.\end{proof}
}\end{proposition}

Similar to Corollary~1 for OMA, we consider a sufficiently large $\hat{\eta}_2$ so that $R_{2,\NOMA}^*(\eta_1,\eta_2)$ in \eqref{Eq:Prop4last} monotonically decreases with $\eta_2$, yielding the following closed-form result.
\begin{corollary}
\linespread{\thmspacing}
(Closed-form maximum average rate, NOMA) \emph{For $\hat{\eta}_2\rightarrow 1$, the objective function in \eqref{Eq:Prop4last} is a monotone decreasing function of $\eta_2$, and we thus obtain $R_{2,\NOMA}^*(\eta_1^*,\eta_2^*)=R_{2,\NOMA}^*(\hat{\eta}_1,\hat{\eta}_2)$ where $\hat{\eta}_1=1-[1/\theta^2 - 1/(1-\hat{\eta}_2)^2]^{-1/2}$.
}\end{corollary}
As shown in Fig.~2 and discussed after Corollary~1, Corollary~2 also holds under $\hat{\eta}_2$ that is much smaller than $1$. We thus consider this as the closed-form approximation, and defer its validation to Sect.~IV.

\section{Numerical Evaluation}
In this section, we numerically validate our analytic results on multiplexing visuo-haptic VR traffic. The default simulation parameters are: $\hat{\eta}_2=0.9$, $c = \alpha/(\alpha-2)$, and $\alpha=4$.

Fig.~3 renders the rate region of the maximized average rate $R_{2,j}^*$ of $\L_2$ and the target rate $\R$ of $\L_1$, under the optimal OMA and NOMA designs. This validates that our closed-form optimal rate expressions $R_{2,j}^*$'s in Corollaries~$1$ and $2$ are only up to $1.75$\% less than the simulated values obtained by using Propositions $2$ and $4$ without the use of Lemma~1. Furthermore, in NOMA, we observe that the performance of the proposed reliability-ordered SIC is significantly degraded when the power-ordered SIC is enforced, calling for an SIC implementation that can decode the weaker signal first if it is more reliable.

Next, compared to OMA, the results in Figs.~3 shows that NOMA performs better for a higher $\R$. In fact, NOMA is capable of achieving a higher rate, since its power-domain multiplexing logarithmically decreases the rate whereas the rate under OMA linearly decreases. In addition, as shown by comparing Figs.~3-a and b, NOMA is also preferable for a higher target visuo-haptic perceptual resolution, i.e. low~$\theta$. In this regime, due to the integrated JND relationship in~\eqref{Eq:JND}, it leads to the higher decoding success probability $\eta_1$ that reduces the rate loss induced by SIC in NOMA, namely, the concatenated decoding at the reception of $\L_2$ signals.

Finally, the maximum required~$\theta$ from which NOMA outperforms OMA for different $\R$'s is specified in Fig.~4. Following the same reasoning as for Fig.~3, a lower $\theta$ with NOMA results in a higher $\eta_1$, thereby minimizing the rate loss induced by the SIC process. With OMA, on the contrary, the higher $\eta_1$ solely decreases the miniblock allocation to $\L_2$, reducing the rate of $\L_2$. For a larger $\R$, the rate of $\L_2$ changes negligibly under the transmit power reduction of NOMA, yet deteriorates significantly under the linear-scale bandwidth reduction of OMA.

\begin{figure}
\centering
\includegraphics[width = \figwidth]{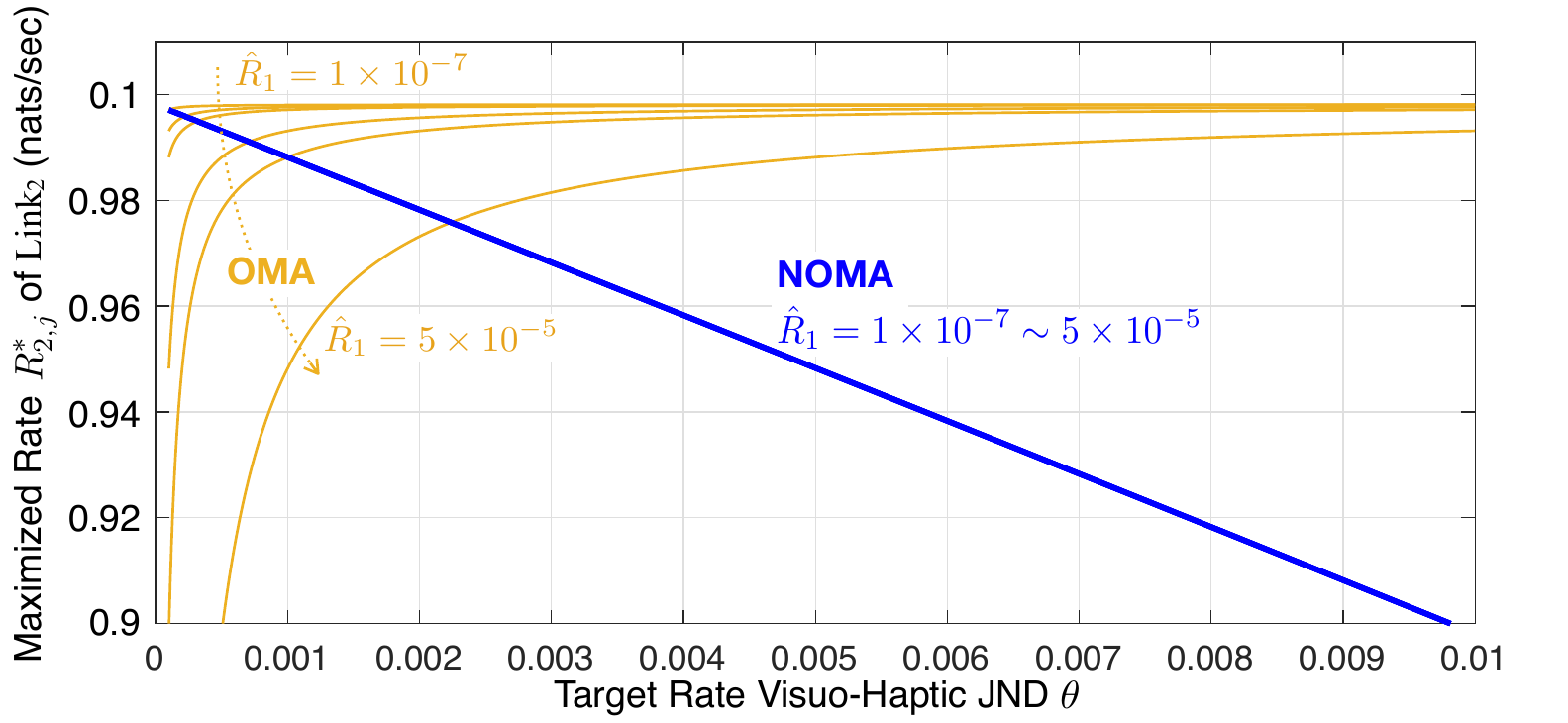}
\caption{\small Maximized average rates $R_{2,\OMA}^*(\eta_2^*)$ and $R_{2,\NOMA}^*(\eta_1^*,\eta_2^*)$ w.r.t. a target visuo-haptic JND threshold $\theta$ for different target rates $\R$'s.}
\end{figure}

\section{Conclusion}
In this paper, we studied the multiplexing design for supporting visuo-haptic VR perceptions through downlink eMBB-URLLC links in cellular systems. Based on our closed-form average rate derivations under OMA and NOMA, we conclude that NOMA with the proposed reliability-ordered SIC outperforms OMA, for a higher target haptic data rate as well as for a higher target visuo-haptic perceptual resolution. A possible extension to this work is to optimize its uplink multiplexing design. Incorporating different types of multimodal perceptions could also be an interesting topic for further research.

\section*{ Acknowledgement}
This paper has benefited from comments and suggestions by P. Popovski and O. Simeone.

\section*{ Appendix - Proofs of Propositions} 
\subsection{ Proof of Proposition 2}
Now that $R_{i,\OMA}(\eta_i)$ monotonically increases with $w_{i,j}$ while decreasing with $w_{\bar{i},j}$ for $\bar{i}\neq i$ with $\bar{i}\in\{1,2\}$, the optimal allocation $w_{1,\OMA}^*$ for $\textsf{P1}$ is determined by the equality constraint \eqref{Eq:Const1}. By applying Proposition~1, \eqref{Eq:Const1} reads

\vspace{-10pt}\small\begin{align}
\eta_1  \log\(1 + G(\eta_1) \) = \R. \label{Eq:Prop2_w1_Exact}
\end{align}\normalsize
Solving this equation yields $w_{1,\OMA}^*$. Applying $w_{2,\OMA}^* = 1- w_{1,\OMA}^*$ and \eqref{Eq:Const3} to $R_{2,\OMA}(\eta_2)$ in Proposition~1 results in $R_{2,\OMA}^*(\eta_2^*)$. The feasible range of $\eta_2$ comes from \eqref{Eq:Const3} and~\eqref{Eq:Const4}. \hfill$\blacksquare$

\subsection{ Proof of Proposition 4}
As $\beta_{1,\NOMA}$ grows, $R_{1,\NOMA}(\eta_1)$ monotonically increases, while $R_{2,\NOMA}(\eta_1,\eta_2)$ decreases. Therefore, the optimal allocation $\beta_{1,\NOMA}^*$ for \textsf{P1} is achieved by the equality constraint \eqref{Eq:Const1} that is given~as
\vspace{-10pt}\small\begin{align}
\eta_1 \log(1 + [\beta_{2,\NOMA}/\beta_{1,\NOMA} + 1/G(\eta_1)]^{-1}) = \R.
\end{align}\normalsize
Solving this equation, we obtain $\beta_{1,\NOMA}^*$. Applying $\beta_{2,\NOMA}^*=1-\beta_{1,\NOMA}^*$ to \eqref{Eq:Const3} to $R_{2,\NOMA}(\eta_1,\eta_2)$ in Proposition~1 results in $R_{2,\NOMA}^*(\eta_1^*,\eta_2^*)$. The feasible range of $\eta_2$ follows from \eqref{Eq:Const3} and~\eqref{Eq:Const4}. \hfill$\blacksquare$

\bibliographystyle{ieeetr}  

\begin{thebibliography}{10}

\bibitem{ABIQualcommVR:17}
{ABI Research and Qualcomm}, ``{Augmented and Virtual Reality: The First Wave
  of 5G Killer Apps},'' {\em White Paper}, Feb. 2017.

\bibitem{EjderVR:17}
{E. Ba{\c s}tu{\u g}, M. Bennis, M. M{\'e}dard, and M. Debbah}, ``{Towards
  Interconnected Virtual Reality: Opportunities, Challenges and Enablers},''
  {\em IEEE Commun. Mag.}, vol.~55, pp.~110--117, Jun. 2017.

\bibitem{ParkWCL:18}
{J. Park, P. Popovski, and O. Simeone}, ``Minimizing latency to support vr
  social interactions over wireless cellular systems via bandwidth
  allocation,'' {\em to appear in IEEE Wireless Commun. Lett. \emph{[Online]}.
  Early access: http://ieeexplore.ieee.org/document/8332500.}

\bibitem{Shi:10}
{Z. Shi, H. Zou, M. Rank, L. Chen, S. Hirche, and H. J. M{\"u}ler}, ``{Effects
  of Packet Loss and Latency on the Temporal Discrimination of Visual-Haptic
  Events},'' {\em IEEE Trans. Haptics}, vol.~3, pp.~28--36, Jan.-Mar. 2010.

\bibitem{UR2Cspaswin:17}
{J. Park, D. Kim, P. Popovski, and S.-L. Kim}, ``{Revisiting Frequency Reuse
  towards Supporting Ultra-Reliable Ubiquitous-Rate Communication},'' {\em
  {Proc. IEEE WiOpt Wksp. SpaSWiN, Paris, France}}, May 2017.

\bibitem{ITU5G:15}
{ITU-R M.2083, IMT Vision -- Framework and Overall Objectives of the Future
  Development of IMT for 2020 and Beyond; Recommendation M.5/BL/22}

\bibitem{Steinbach:12}
E.~Steinbach, S.~Hirche, M.~Ernst, F.~Brandi, R.~Chaudhari, J.~Kammerl, and
  I.~Vittorias, ``{Haptic Communications},'' {\em {Proc. IEEE}}, vol.~100,
  pp.~937--956, Apr. 2012.

\bibitem{Zhang:18}
{Q. Zhang, J. Liu, and G. Zhao}, ``{Towards 5G Enabled Tactile Robotic
  Telesurgery},'' {\em \emph{[Online]}. ArXiv preprint:
  https://arxiv.org/abs/1803.03586.}

\bibitem{PetarURLLC:17}
{P. Popovski, J. J. Nielsen, C. Stefanovic, E. de Carvalho, E. G. Str\"{o}m, K.
  F. Trillingsgaard, A. Bana, D. Kim, R. Kotaba, J. Park, and R. B.
  S\o{}rensen}, ``{Wireless Access for Ultra-Reliable Low-Latency Communication
  (URLLC): Principles and Building Blocks},'' {\em IEEE Netw.}, vol.~32,
  pp.~16--23, Mar. 2018.

\bibitem{MehdiURLLC:18}
{M. Bennis, M. Debbah, and V. Poor}, ``{Ultra-Reliable and Low-Latency Wireless
  Communication: Tail, Risk and Scale},'' {\em \emph{[Online]}. ArXiv preprint:
  https://arxiv.org/abs/1801.01270.}

\bibitem{Ernst:2002aa}
M.~O. Ernst and M.~S. Banks, ``{Humans Integrate Visual and Haptic Information
  in a Statistically Optimal Fashion},'' {\em Nature}, vol.~415, pp.~429--433,
  Jan. 2002.

\bibitem{ShiHirche:16}
{M. Rank, Z. Shi, H. J. M{\"u}ller, and S. Hirche}, ``{Predictive Communication
  Quality Control in Haptic Teleoperation With Time Delay and Packet Loss},''
  {\em IEEE Trans. Hum. Mach. Syst.}, vol.~46, pp.~581--592, Aug. 2016.

\bibitem{3GPPMUST:2015}
{3GPP TR 36.859, Study on Downlink Multiuser Superposition Transmission (MUST)
  for LTE.}

\bibitem{OsvaldoAR:17}
A.~AL-Shuwaili and O.~Simeone, ``{Energy-Efficient Resource Allocation for
  Mobile Edge Computing-Based Augmented Reality Applications},'' {\em IEEE
  Wireless Commun. Lett.}, vol.~6, pp.~398--401, Apr. 2017.

\bibitem{ChenSaad:17}
M.~Chen, W.~Saad, and C.~Yin, ``{Resource Management for Wireless Virtual
  Reality: Machine Learning Meets Multi-Attribute Utility},'' {\em Proc. IEEE
  GLOBECOM, Singapore}, Dec. 2017.

\bibitem{Elbamby:18}
{M. S. Elbamby, C. Perfecto, M. Bennis, and K. Doppler}, ``{Toward Low-Latency
  and Ultra-Reliable Virtual Reality},'' {\em IEEE Netw.}, vol.~32, pp.~78--84,
  Mar. 2018.

\bibitem{Petar5G:18}
{P. Popovski, K. F. Trillingsgaard, O. Simeone, and G. Durisi}, ``{5G Wireless
  Network Slicing for eMBB, URLLC, and mMTC: A Communication-Theoretic View},''
  {\em \emph{[Online]}. ArXiv preprint: https://arxiv.org/abs/1804.05057.}

\bibitem{Andrews:2011bg}
J.~G. Andrews, F.~Baccelli, and R.~K. Ganti, ``{A Tractable Approach to
  Coverage and Rate in Cellular Networks},'' {\em IEEE Trans. Commun.},
  vol.~59, pp.~3122--3134, Nov. 2011.

\bibitem{HaenggiSG}
M.~Haenggi, {\em {Stochastic Geometry for Wireless Networks}}.
\newblock Cambridge Univ. Press, 2013.

\bibitem{JHParkTWC:15}
J.~Park, S.-L. Kim, and J.~Zander, ``{Tractable Resource Management with Uplink
  Decoupled Millimeter-Wave Overlay in Ultra-Dense Cellular Networks},'' {\em
  IEEE Trans. Wireless Commun.}, vol.~15, pp.~4362--4379, Jun. 2016.

\bibitem{TseBook:FundamaentalsWC:2005}
D.~N.~C. Tse and P.~Viswanath, {\em {Fundamentals of Wireless Communications}}.
\newblock Cambridge University Press, 2005.

\bibitem{TseOC:07}
{A. S. Avestimehr and D. N. C. Tse}, ``{Outage Capacity of the Fading Relay
  Channel in the Low-SNR Regime},'' {\em IEEE Trans. Inf. Theory}, vol.~53,
  pp.~1401--1415, Apr. 2007.

\bibitem{DurisiPolyanski:14}
{W. Yang, G. Durisi, T. Koch, and Y. Polyanski}, ``{Quasi-Static
  Multiple-Antenna Fading Channels at Finite Blocklength},'' {\em IEEE Trans.
  Inf. Theory}, vol.~60, pp.~4232--4265, Jul. 2014.

\bibitem{JindalSIC:09}
{J. Blomer and N. Jindal}, ``{Transmission Capacity of Wireless Ad Hoc
  Networks: Successive Interference Cancellation vs. Joint Detection},'' {\em
  Proc. IEEE ICC, Dresden, Germany}, Jun. 2009.

\bibitem{Haenggi:ISIT14}
X.~Zhang and M.~Haenggi, ``{A Stochastic Geometry Analysis of Inter-Cell
  Interference Coordination and Intra-Cell Diversity},'' {\em IEEE Trans.
  Wireless Commun.}, vol.~13, pp.~6655--6669, Dec. 2014.

\bibitem{RelCovWCL:17}
J.~Park and P.~Popovski, ``{Coverage and Rate of Downlink Sequence
  Transmissions with Reliability Guarantees},'' {\em IEEE Wireless Commun.
  Lett.}, vol.~6, pp.~722--725, Aug. 2017.

\end{thebibliography}

\end{document}